\documentclass[a4paper,11pt,onecolumn] {article}
\usepackage{geometry}
\usepackage{graphicx}
\usepackage{amssymb}
\usepackage{epstopdf}
\usepackage{amsmath}
\date{}

\begin{document}
\title{Comparing domain wall synapse with other Non Volatile Memory devices for on-chip learning in Analog Hardware Neural Network}
\author{Divya Kaushik$^{*1}$, Utkarsh Singh$^{*2}$, Upasana Sahu$^{1}$, \\ Indu Sreedevi$^{2}$ and Debanjan Bhowmik$^{1}$ \\ $^*$These authors contributed equally to the work. \\ $^1$Department of Electrical Engineering, \\ Indian Institute of Technology Delhi, New Delhi-110016, India \\ $^2$Department of Electronics and Communication Engineering, \\ Delhi Technological University, Delhi-110042, India \\ E-mail: debanjan@ee.iitd.ac.in}
\maketitle

\begin{abstract}
Resistive Random Access Memory (RRAM) and Phase Change Memory (PCM) devices have been popularly used as synapses in crossbar array based analog Neural Network (NN) circuit to achieve more energy and time efficient data classification compared to conventional computers. Here we demonstrate the advantages of recently proposed spin orbit torque driven Domain Wall (DW) device as synapse compared to the RRAM and PCM devices with respect to on-chip learning (training in hardware) in such NN. Synaptic characteristic of DW synapse, obtained by us from micromagnetic modeling, turns out to be much more linear and symmetric (between positive and negative update) than that of RRAM and PCM synapse. This makes design of peripheral analog circuits for on-chip learning much easier in DW synapse based NN compared to that for RRAM and PCM synapses. We next incorporate the DW synapse as a Verilog-A model in the crossbar array based  NN circuit we design on SPICE circuit simulator. Successful on-chip learning is demonstrated through SPICE simulations on the popular Fisher's Iris dataset. Time and energy required for learning turn out to be orders of magnitude lower for DW synapse based NN circuit compared to that for RRAM and PCM synapse based NN circuits.
\end{abstract}


\section{\label{sec:level1}Introduction}


Crossbar array based analog hardware Neural Network (NN) is considered to be extremely time and energy efficient in executing NN algorithms for data classification applications because it computes at the location of memory itself unlike CPU, GPU and even the recent digital neuromorphic chips which all have memory and computing separate at their smallest cores \cite{ GeffBurrJPhysDReview,PCMReview_AbuSebastian,Kaushik_IEEEReview,IBMTrueNorth,Loihi_Intel,GeffBurrIEDM2015}. Such crossbar based NN needs an analog Non Volatile Memory (NVM) device, also known as synapse, at each of the intersection points of the crossbars. Typically a Resistive Random Access Memory (RRAM) or a Phase Change Memory (PCM) device is used as synapse \cite{GeffBurrJPhysDReview,PCMNature, memristorNature, PCMManan,MSuriBook}. Training the NN in hardware (on-chip learning) is achieved by modulating the conductances of the synapses, corresponding to weights stored in synapses, with electrical programming pulses at every iteration. Though the conductance of RRAM and PCM synapses changes by orders of magnitude due to programming pulses, conductance response characteristic is highly non-linear and asymmetric (between positive and negative conductance update) \cite{GeffBurrJPhysDReview,GeffBurrTED,ShimengIEDM,RPUFrontNeuroscience}. This leads to issues with design of peripheral circuits for on-chip learning. Learning accuracy suffers. Time and energy consumed in the learning process are also very high \cite{GeffBurrJPhysDReview,GeffBurrTED,ShimengIEDM,RPUFrontNeuroscience,PCMManan}. 

Spin orbit torque driven Domain Wall (DW) device based on heavy metal-ferromagnet hetero-structure has been recently proposed and experimentally demonstrated to exhibit synaptic behaviour \cite{Kaushik_IEEEReview,Kaushik_BioMedCircuit,Saxena,LongYouDWSynapseExpt,Bhowmik_JMMM, JeanAnnDomainWallNeuron1, Parker_spintronicspiking,SkyrmionSynapse}. In Section II of this paper, we simulate such DW synapse based on experimentally calibrated micromagnetic model. We show that though the range of conductance variation is much smaller for DW synapse than RRAM and PCM synapse, the conductance response of DW synapse to programming current pulse is linear and symmetric unlike RRAM and PCM synapse. In Section III, we design crossbar array of DW synapses in SPICE circuit simulator, with the synapses being Verilog-A models developed from our micromagnetic simulation results. Fully Connected Neural Network (FCNN) algorithm, with Stochastic Gradient Descent (SGD) based weight/ conductance update, has been used here for on-chip learning \cite{Bhowmik_JMMM,LeCun}. Conductance of DW synapse has been quantized here unlike in Bhowmik \textit{et al.} \cite{Bhowmik_JMMM} to take the effect of DW pinning by defects into account \cite{DWPinning1,DWPinning2,DWPinning3}. Despite the quantization, high accuracy is obtained on a popular machine learning dataset- Fisher's Iris \cite{Fisher}, in our circuit simulations. We next show that the time taken and energy consumed for on-chip learning of the DW synapse based NN circuit are orders of magnitude lower than RRAM and PCM synapse based NN circuit. 
Section IV concludes the paper. To the best of our knowledge, this is the first comparison study between a spintronic synapse and RRAM/ PCM synapse, with respect to on-chip learning in NN hardware. 
\section{Device Level Comparison}
\begin{figure}
    \centering
    \includegraphics[width=0.5\textwidth]{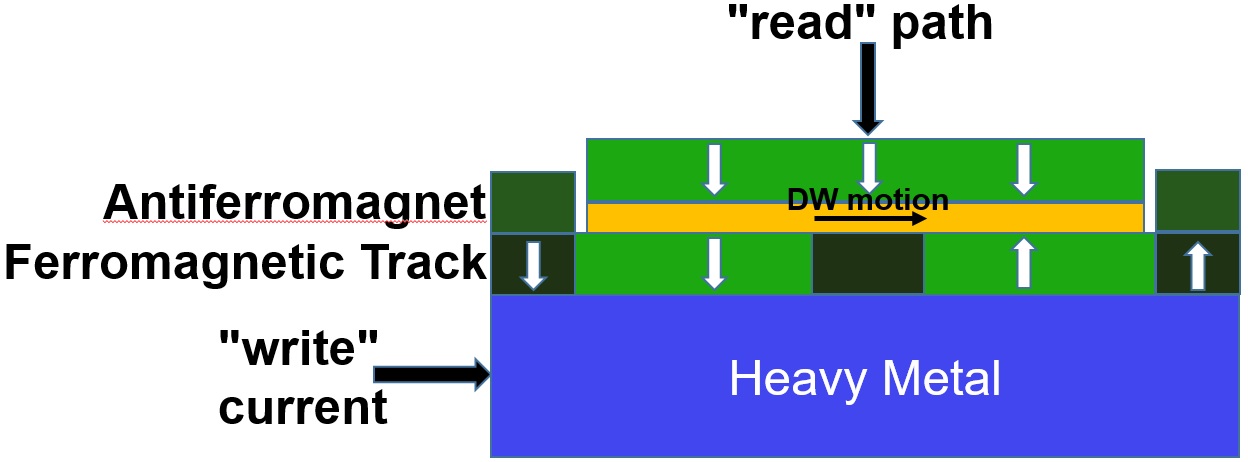}
    \caption{Schematic of Domain Wall (DW) synapse.} 
    \label{Fig.1}
\end{figure}

\begin{figure}
    \centering
    \includegraphics[width=0.6\textwidth]{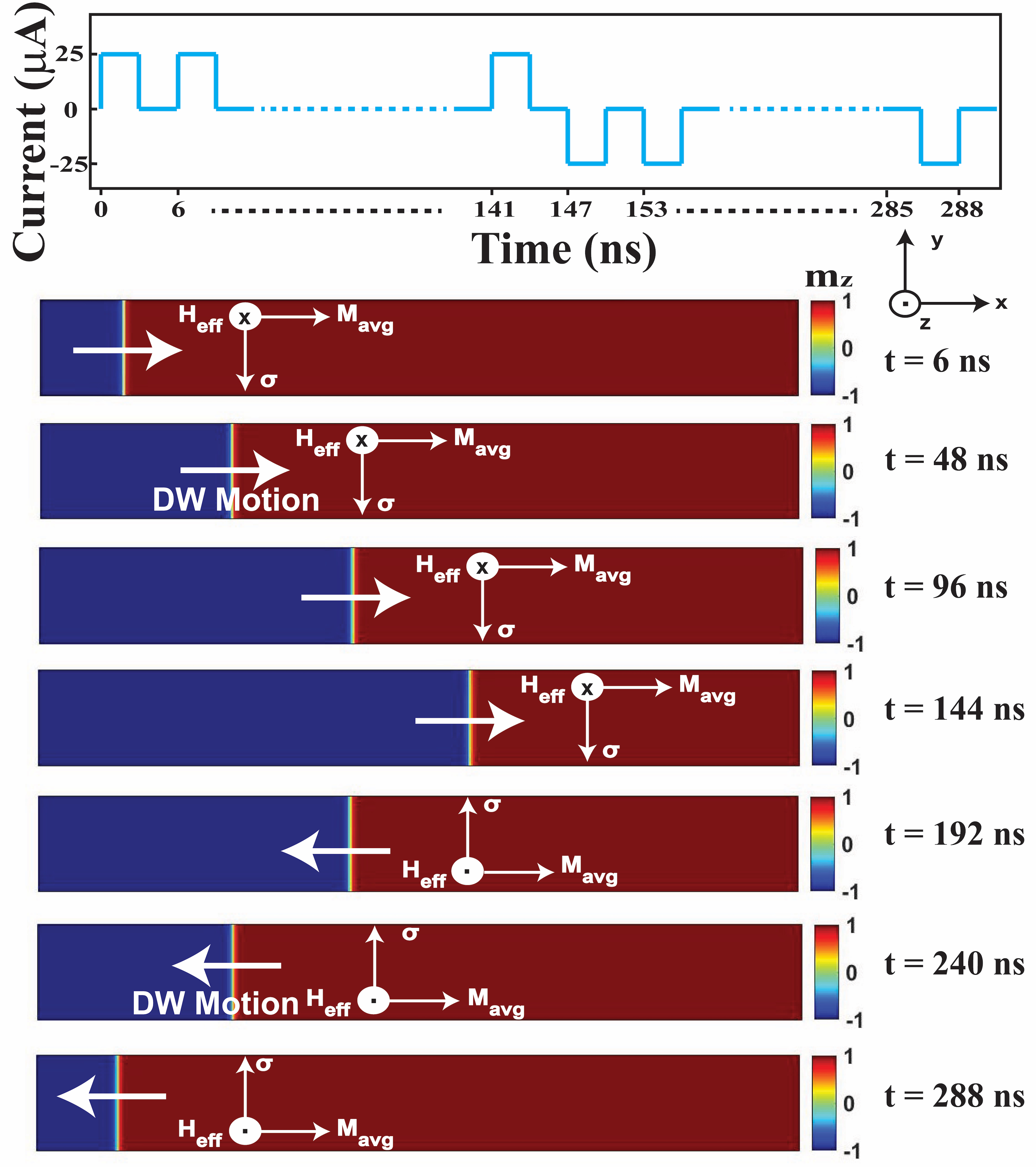}
    \caption{(a) "Write" current pulses applied on the heavy metal layer of constant magnitude (25 $\mu$A) and opposite polarities as a function of time. (b) DW motion in the ferromagnetic layer, above the heavy metal layer, shown for different time instants corresponding to different current pulses in (a).} 
    \label{Fig.2}
\end{figure}

Schematic of our heavy metal/ ferromagnetic metal hetero-structure based domain wall based synapse is shown in Fig. ~\ref{Fig.1}. The operating physics of the device has been discussed extensively in \cite{Kaushik_IEEEReview,Kaushik_BioMedCircuit,Saxena,Bhowmik_JMMM}. The core physics is that of spin orbit torque driven DW motion, which has been extensively studied through simulations and experiments in the past\cite{Emori,Ryu,Bhowmik,Miron}. When in-plane current ("write" current) flows through the heavy metal layer ("write" path), a DW in the ferromagnetic layer above the it experiences spin orbit torque. If the DW is of Neel type due to presence of Dzyaloshinskii Moriya Interaction (DMI) \cite{Emori,Ryu,Emori_2,Sampaio} at the interface, it moves even in the absence of magnetic field,as observed in several experiments\cite{Emori,Ryu,Emori_2} and also our micromagnetic simulations (Fig.~\ref{Fig.2}).
\begin{figure}[!t]
    \centering
    \includegraphics[width=0.6\textwidth]{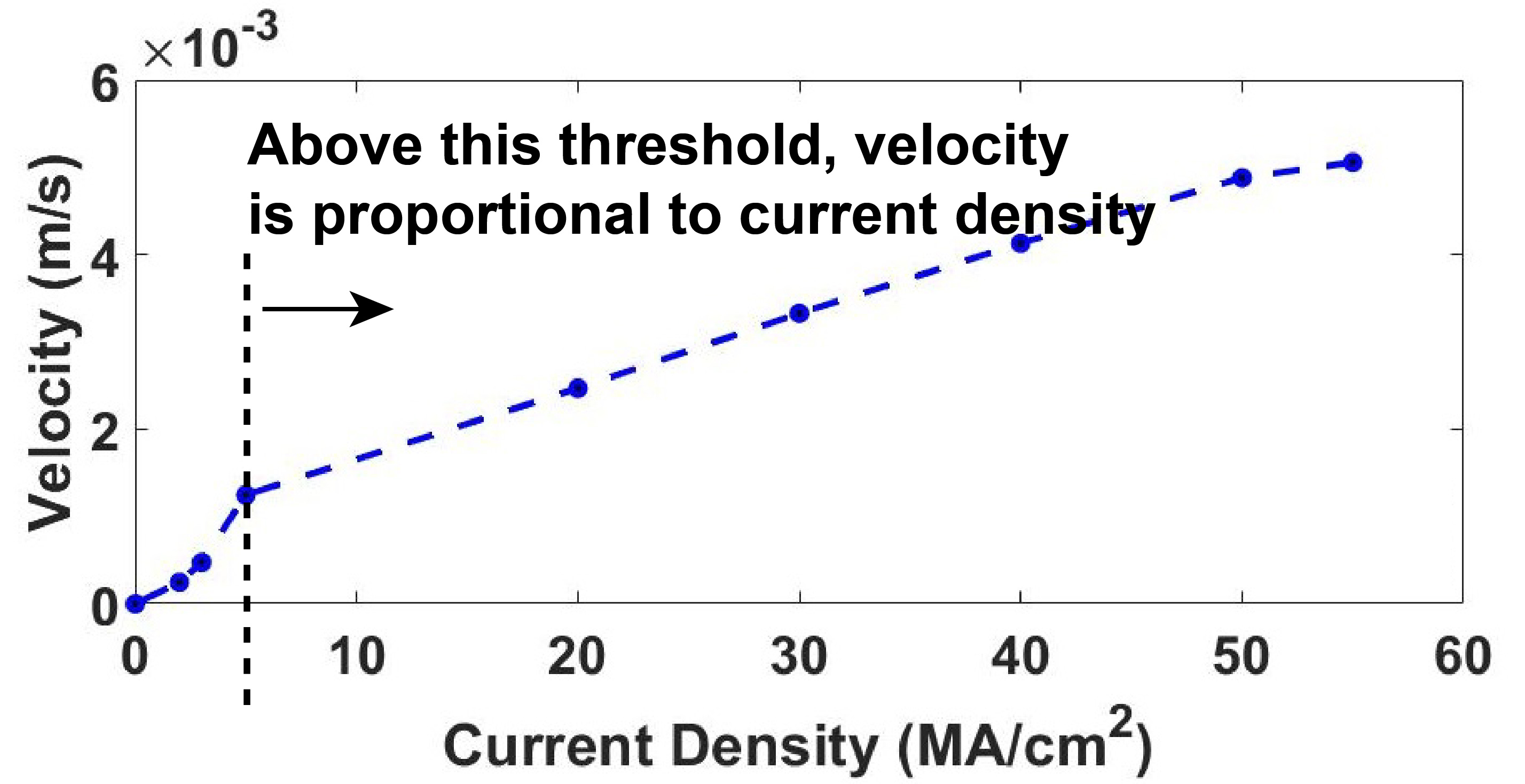}
    {\includegraphics[width=0.59\textwidth]{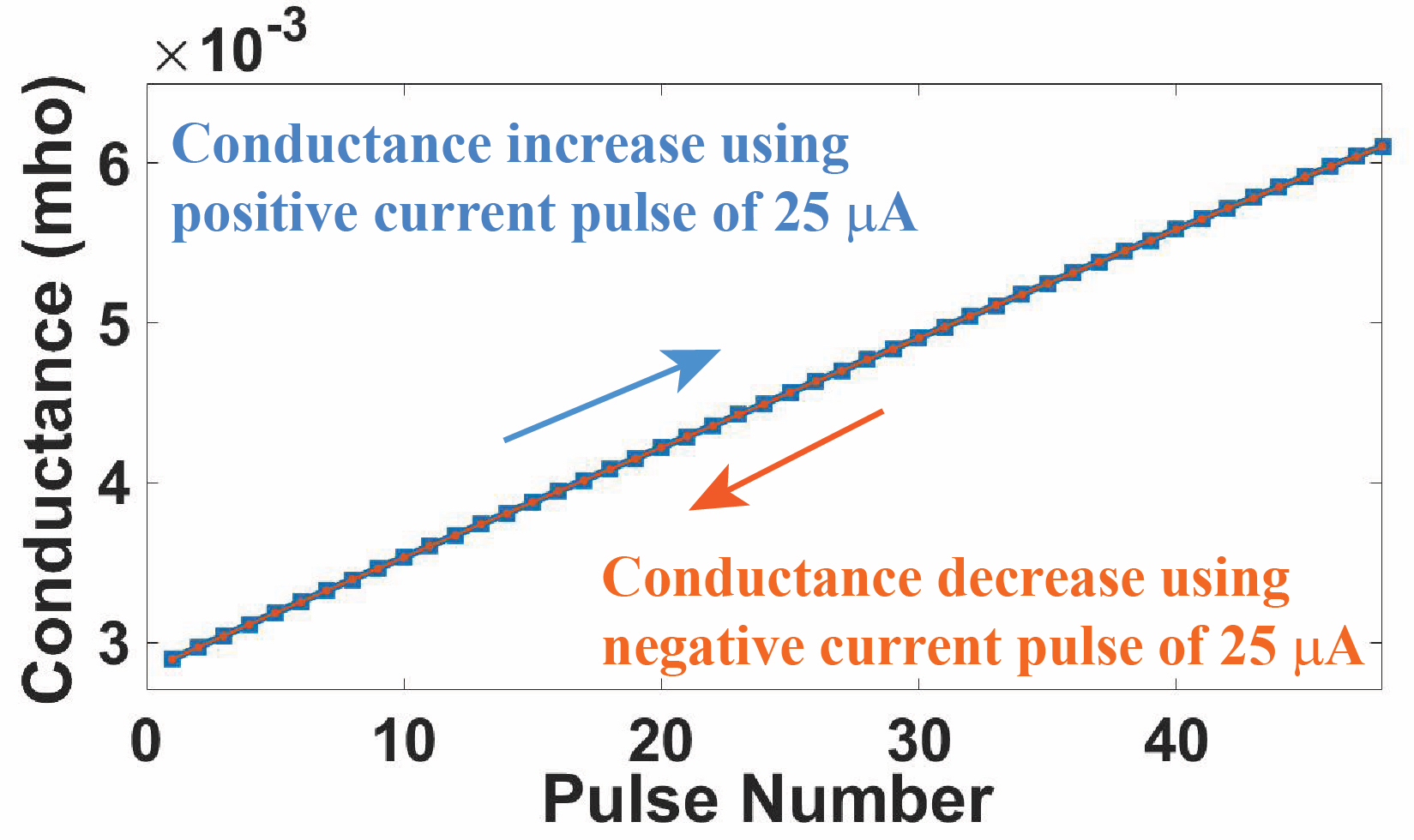}}
    \caption{(a) Velocity vs. in-plane charge current density, obtained from micro-magnetic simulation of the ferromagnetic free layer of the DW synapse device in which the DW moves. (b) Conductance response of DW synapse as a function of programming "write" current pulse.}
    \label{Fig.3}
\end{figure}

In this paper, we consider a device with lateral dimensions 1000 nm $\times$ 50 nm. Thickness of the heavy metal (Pt) layer is taken to be 10 nm, which is greater than the spin diffusion length. Hence, we can consider the vertical spin current density injected by the heavy metal layer on the ferromagnetic layer above it ($J_s$) = in-plane charge current density ($J_c$) $\times$  spin Hall angle (0.07 here, considering Pt)  \cite{Liu_Science, Liu_PRL1, Liu_PRL2}. Thickness of the  ferromagnetic layer above the heavy metal layer is taken to be 1 nm. Dynamics of the moments of this layer under the influence of this spin current is simulated using micromagnetic simulation package "mumax3" \cite{mumax} to model such spin current driven DW motion inside it. We choose micro-magnetic simulation parameters for the ferromagnetic layer based on that used for Pt(heavy metal)/ CoFe (ferromagnet)/ MgO devices in the simulation study of Emori \textit{et al.} \cite{Emori_2}, which is based on experimentally observed spin orbit torque driven DW motion in the same devices. The parameters can also be found in Supplementary Material (Section 1) accompanying this paper.

Since the DW is of Neel type (DMI = $1.2 \times 10^{-3}$ J/m$^2$), average magnetization inside the wall ($\vec{M_{avg}}$) and direction of spin polarization of the electrons at the interface of heavy metal and ferromagnet due to current flowing through heavy metal ($\vec{\sigma}$) form a non-zero cross product (Fig.~\ref{Fig.2}). The effective magnetic field experienced by DW is equal to that cross-product \cite{Emori,Bhowmik,Emori_2}. As a result, DW moves as seen in our micro-magnetic simulations (Fig.~\ref{Fig.2}). Triangular notch regions with Perpendicular Magnetic Anisotropy (PMA) constant = $9 \times 10^{5}$ J/m$^3$ are present on the edges of the simulated ferromagnetic layer in our simulations. PMA in rest of the layer = $8 \times 10^{5}$ J/m$^3$. These notch regions mimic defects, which pin the domain wall for in-plane charge current lower than a certain threshold value \cite{Saxena,DWPinning1,DWPinning2,DWPinning3,Sampaio}. Hence, our micro-magnetic simulation Fig.(~\ref{Fig.3}(a)) shows that only above a certain threshold value of current density ($\approx 5\times10^{6}$ A/cm$^2$), velocity of the domain wall is linearly proportional to the current density \cite{Martinez_2}. Hence in our device we have only moved the domain wall with a current pulse (3 ns long) of fixed magnitude (25 $\mu$A)(Fig.~\ref{Fig.2}),corresponding to a current density of $5\times10^{6}$ A/cm$^2$  (Fig.~\ref{Fig.2}) so that the domain wall is never pinned by defects \cite{DWPinning1,DWPinning2,DWPinning3}. Pinned ferromagnetic regions are present at each edge of the free layer to stabilize the DW at the edge and prevent it from getting destroyed\cite{Kaushik_IEEEReview,Kaushik_BioMedCircuit, Sokalski}.

\begin{figure}[!t]
    \centering
    {\includegraphics[width=0.6\textwidth]{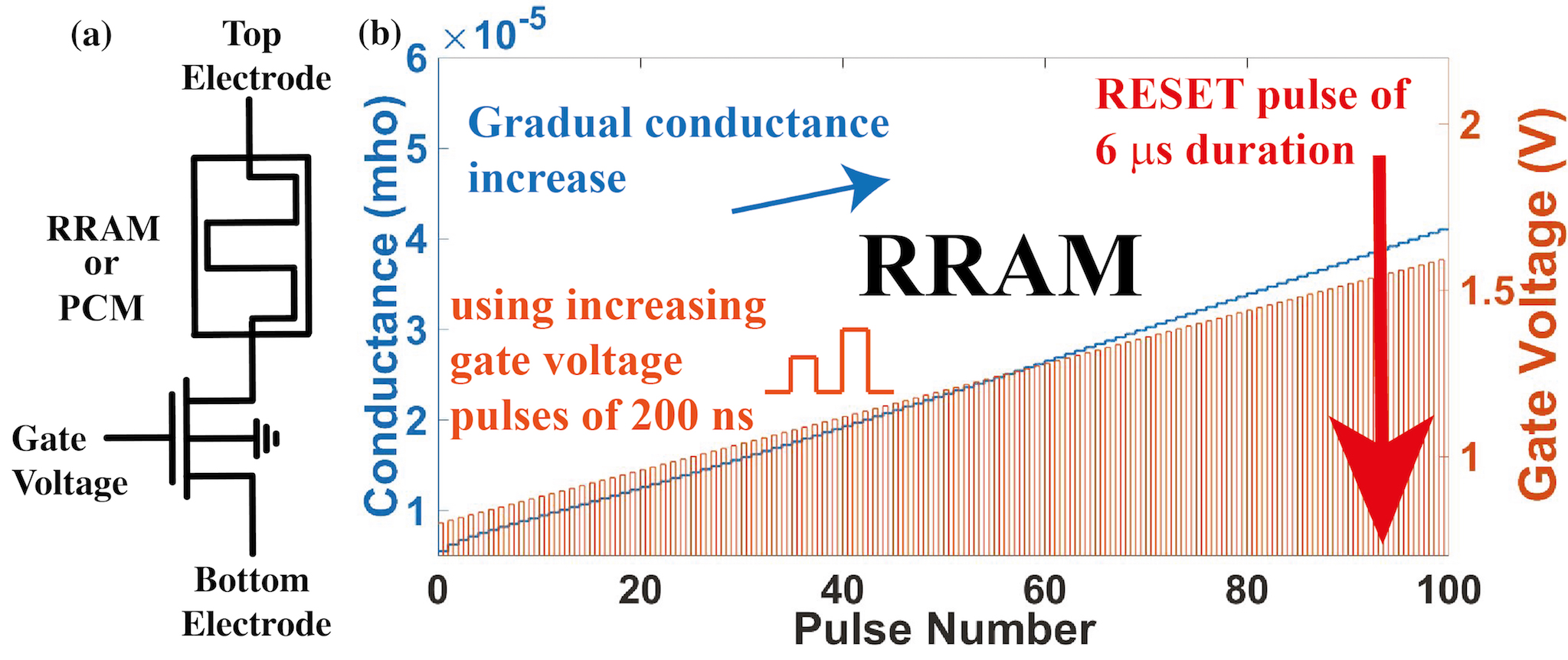}}
    {\includegraphics[width=0.6\textwidth]{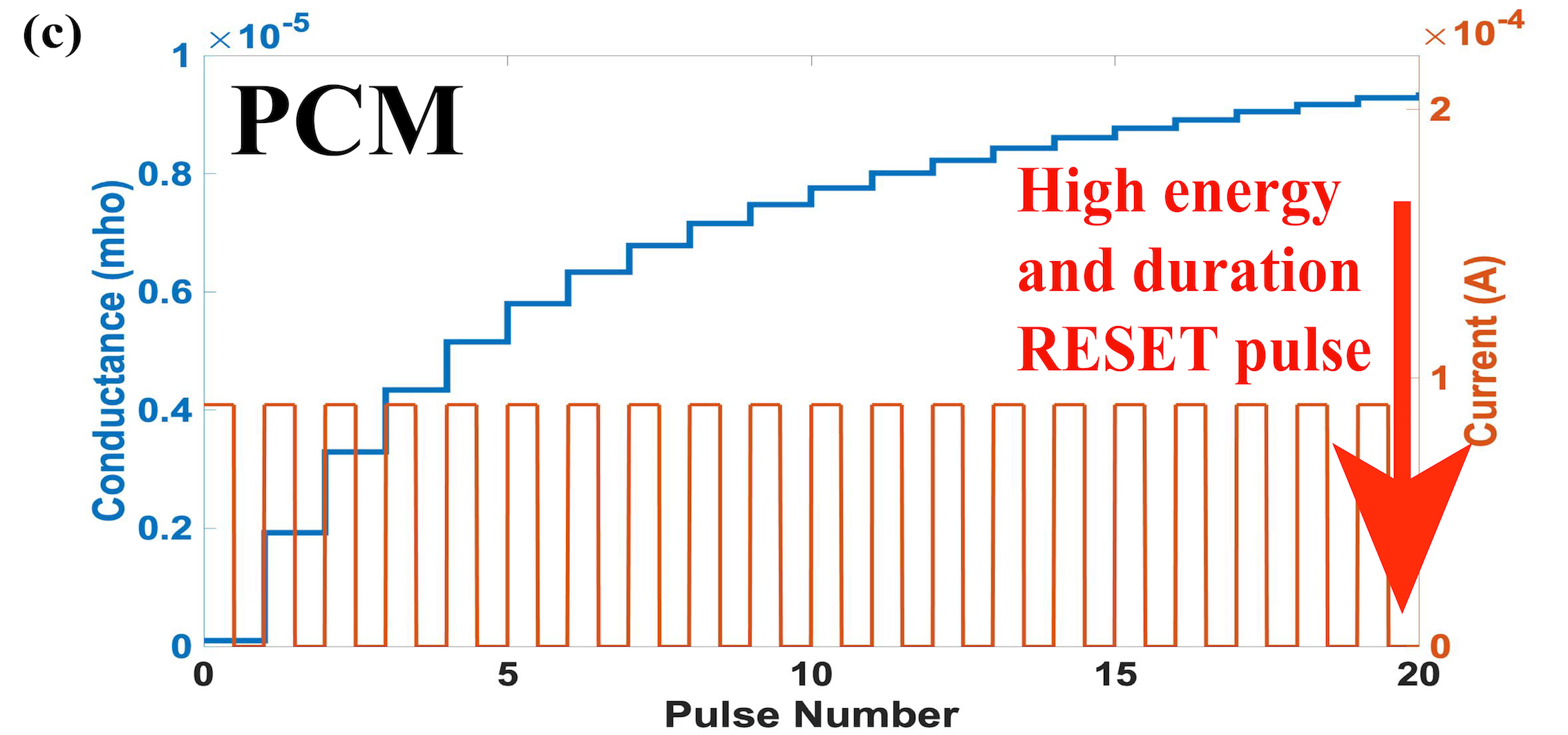}}

    \caption{(a) Schematic of RRAM device connected to transistor. (b) Conductance increase in RRAM by applying gate voltage pulses of increasing magnitude at the transistor of in  (c) Conductance increase in PCM device by application of constant magnitude gate voltage pulses at the transistor.}

    \label{Fig5}
\end{figure}

Following \cite{Kaushik_BioMedCircuit} conductance of the "read" path (vertical tunnel junction structure in Fig. ~\ref{Fig.1}) of the synapse is given by,
\begin{equation}
\emph{ $G^{synapse}$ =  ($G_{max}$ + $G_{min}$ )/2 - (($G_{max}$ - $G_{min}$ )$<m_{z}>$/2})
\end{equation} where $<m_{z}>$ represents the average out of plane magnetization component of the free layer ($<m_{z}>$ = 1 corresponds to up and $<m_{z}>$ = -1 corresponds to down), $G_{max}$ is maximum conductance of MTJ and $G_{min}$ is minimum conductance of MTJ. Taking the Resistance-Area product  of the MTJ to be \cite{Zhu} $4.04\times10^{-12}$ ohm/m$^2$ and TMR ratio of 120 $\%$ \cite{Ikeda},    $G_{min}$  $\approx2.9\times10^{-3}$ mho and $G_{max}$  $\approx6.1\times10^{-3}$ mho. Moment of the fixed layer is in down direction (Fig. ~\ref{Fig.1}).

As observed from our micromagnetic simulation, "write" current pulse of magnitude 25 $\mu$A and positive polarity always moves DW to the right by a fixed distance of ~$\approx$ 20 nm  (Fig. ~\ref{Fig.2}). Hence $<m_{z}>$ decreases and following equation (1) conductance increases by a step of $0.071 \times 10^{-3}$ mho (Fig. ~\ref{Fig.3}(b)). Current pulse of same magnitude and negative polarity moves DW to the left, $<m_{z}>$ increases and conductance decreases by the same step of $0.071 \times 10^{-3}$ mho (Fig. ~\ref{Fig.3}(b)). Hence conductance response to a series of programming "write" current pulses of equal magnitude (25 $\mu$A) is linear and is also symmetric between  positive and negative pulses. Also conductance of DW synapse and hence the corresponding weight of the synapse only takes quantized values and thus we take defect pinning into account. Energy consumed through Joule heating per programming pulse of 25 $\mu$A  for conductance increase/ decrease by a single step is calculated to be 0.18 fJ (Table I).

\begin{figure}[!t]
    \centering
    \includegraphics[width=1\textwidth]{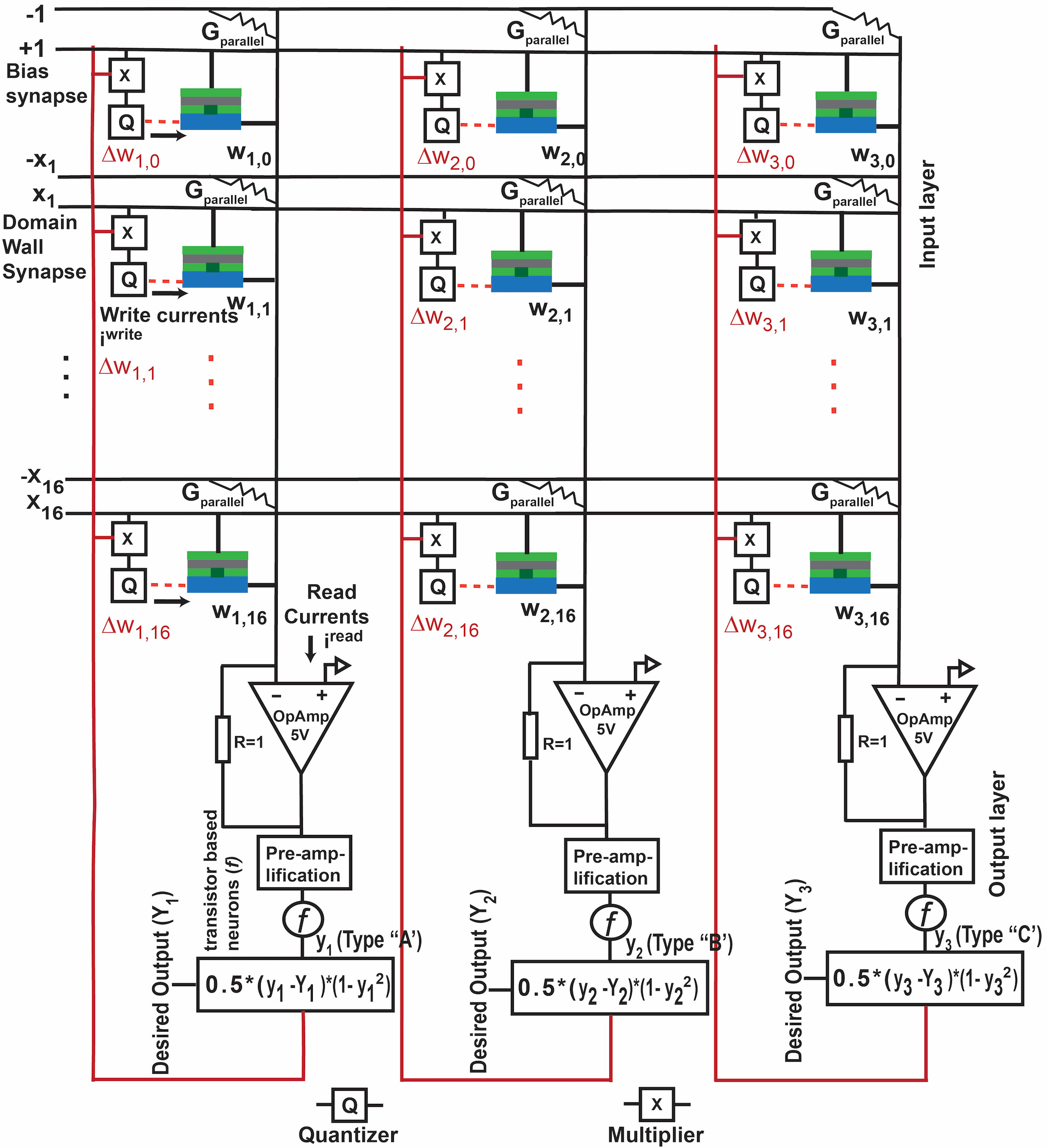}
    \caption{Schematic of designed DW synapse based crossbar array along with analog peripheral circuits for on-chip learning}
    \label{Fig.4}
\end{figure}

Next we compare the conductance response of this DW synapse with that of typical RRAM and PCM synapse. Verilog-A model provided by \cite{Shimeng}, experimentally benchmarked against \cite{ExperimentalMemristor}, has been used for RRAM modeling. Following the 1T1M (one transistor, one memristor) configuration \cite{Miao, Jiang, Li} we connect this RRAM device with a 65 nm technology node transistor (from UMC library) in Cadence Virtuoso circuit simulator (Fig.~\ref{Fig5}(a)). We observe that when gate voltage pulses of fixed magnitude  and duration (200 ns) are applied at the gate of the transistor for conductance increase (voltage of top electrode kept higher than that of bottom electrode for that purpose) (Check Supplementary Material- Section 2), the conductance of the RRAM synapse does not go up linearly unlike the domain wall synapse. In fact the conductance just saturates to a fixed value (Check Supplementary Material- Section 2). To achieve a linear increase in conductance gate voltage pulses of increasing magnitude (SET pulses) need to be applied (Fig.~\ref{Fig5}(b)). This has been observed experimentally in the RRAM devices of \cite {Miao,Li,Woo}. Thus, though the conductance varies over a much wider range for RRAM synapse than DW synapse (Table I), the conductance response is inherently non-linear in nature. As a result, if a certain value of weight update is needed for any synapse for an iteration during on-chip, different magnitude of voltage pulses may need to be applied to bring about the same weight update, depending on what weight/ conductance value of the RRAM synapse is before that iteration. This makes designing the analog peripheral circuit for weight update very complicated. In fact, the demonstrations of on-chip learning in RRAM based crossbar NN array so far use a digital FPGA unit or an on-chip CMOS based digital processor, connected to the analog crossbar array, for weight update\cite{Li,UMichNatureElectronics}. ADC-s and DAC-s needed as a result, which can potentially consume a lot of energy and slow down the circuit. Energy consumed in the 1T1M circuit of Fig.~\ref{Fig5}(a) ranges between 12 pJ (minimum gate voltage) and 51 pJ (maximum gate voltage), which is much larger than energy consumed for weight/ conductance update by a single step in a domain wall synapse (Fig.~\ref{Fig.3}(b)) (Table I). Apart from non-linearity, another issue with conductance response of RRAM synapse is asymmetry between positive and negative update of conductance. If we apply the same gate voltage pulses as in Fig.~\ref{Fig5}(b) in the reverse order in order to decrease the conductance of the synapse (bottom electrode at higher voltage than top for that purpose), we see that the conductance hardly decreases (Check Supplementary Material- Section 2). Rather in order to decrease conductance by a certain step, long duration (6 $\mu$s) and high magnitude (2.5 V) voltage pulse (hence high energy consuming), known as RESET pulse, needs to be applied at the gate of the transistor for abrupt conductance decrease to the smallest value. It is followed by pulses of gradually increasing voltage pulses (SET pulses) to then increase the conductance. 
\begin{center}
\begin{table*}[hbt!]
\scalebox{0.6}{%
\begin{tabular}{|c||c|c|c|c|c|c|c|} 

 \hline
  &  & \textbf{Number}&\textbf{Energy} & \textbf{Energy} & \textbf{Train} & \textbf{Test} & \textbf{Total energy}\\
\textbf{Synaptic} & \textbf{Conductance}& \textbf{of} & \textbf{per pulse}  &\textbf{per pulse}  &  \textbf{accuracy} & \textbf{accuracy}  & \textbf{consumed in}\\
  \textbf{devices} & \textbf{range (mho)}& \textbf{conductance} & \textbf{(conductance}  &\textbf{(conductance}  &  \textbf{for on-chip} & \textbf{for on-chip}  & \textbf{all synapses for}\\
  
   & & \textbf{states} & \textbf{increase)}  &\textbf{decrease)}  &  \textbf{learning ($\%$)} & \textbf{learning ($\%$)}  & \textbf{on-chip learning}\\
 \hline
 Domain Wall  & 2.9m - 6.1m & 48 & 0.18 fJ & 0.18 fJ  & 89  & 90 & 9 fJ  \\ 
 \hline
 RRAM  & 3$\mu$ - 30$\mu$ & 100 & 12 pJ - 51 pJ & 2.28 nJ (abrupt Reset) & 93 & 94 & 1 $\mu$J  \\ 
 \hline
 PCM & 0.1$\mu$ - 9.3 $\mu$ & 20 & 5 pJ & 30 pJ (abrupt Reset)  & 89 & 92 & 1.1 $\mu$J  \\
\hline
\end{tabular}}
\caption{\label{tab:table1}Performance comparison between Domain Wall, RRAM and PCM based fully connected neural networks for on-chip learning.}
\end{table*}
\end{center}

Conductance response characteristic of PCM synapse we simulated, based on model developed in Nandakumar \textit{et al.} \cite{Nandakumar} (See Supplementary Material- Section 3 for more details), is more linear than RRAM i.e. programming current pulse of fixed magnitude 90 $\mu A$ and duration 50 ns increase the conductance of the PCM synapse fairly linearly for a larger number of pulses (~$\approx$ 12) (Fig.~\ref{Fig5}(c)). Energy associated with each such pulse is 5 pJ \cite{Nandakumar,PCMReview_AbuSebastian}, still much higher than that for domain wall synapse (Table I). Conductance decrease on the other hand is carried out by an abrupt RESET pulse that consumes 30pJ energy each \cite{PCMReview_AbuSebastian}, followed by a series of SET pulses much like RRAM synapse. Thus the conductance response characteristic of PCM synapse is still asymmetric like RRAM synapse.

\section{Network Level Comparison}
Next we design crossbar array based Fully Connected Neural Network (FCNN) with domain wall synapses \cite{Bhowmik_JMMM} and compare the energy and speed performance for on-chip learning with that for equivalent FCNN designed with RRAM and PCM synapses. It is to be noted that this NN is of the second generation non-spiking type \cite{generation} and uses standard Stochastic Gradient Descent (SGD) algorithm for weight update \cite{LeCun}. Verilog-A model of domain wall synapse is designed, based on its conductance response obtained from micromagnetic physics as shown in Fig.~\ref{Fig.3}(b)) and inserted in crossbar schematic designed on Cadence Virtuoso circuit simulator (Fig.~\ref{Fig.4}). 


Fisher's Iris dataset, a popular machine learning dataset, is used for the training \cite{Fisher}. Since the dataset is not completely linearly separable, in order to carry out accurate classification on it with a FCNN without a hidden layer which we design here, the 4 input features corresponding to each sample are passed through some basic filters first to convert to 16 features \cite{Udayan}. Input voltages, proportional to these 16 input features, are applied on the crossbar as shown in (Fig.~\ref{Fig.4} ). Read currents, proportional to product of weight of the synapse and each input feature, add up following Kirchhoff's current law and enter the neuron/ activation function circuit at each output node. Thus the input Vector- weight Matrix Multiplication (VMM) is carried out in the crossbar array. \cite{GeffBurrJPhysDReview,Bhowmik_JMMM}."tanh"  neuron/activation function ($f$) acts on the read current at each output node. This function has been designed with transistors in differential amplifier configuration, as shown in Bhowmik \textit{et al.} \cite{Bhowmik_JMMM}. A weight update circuit follows which calculates the common part of weight update at each output node, using the same SGD method and circuit described in Bhowmik \textit{et al.}  \cite{Bhowmik_JMMM}. The common part of weight update computed at each output node is next multiplied with the inputs using the multiplier circuit (x) as shown in(Fig.~\ref{Fig.4}). In Bhowmik \textit{et al.} \cite{Bhowmik_JMMM}, write current proportional to the output of the multiplier (x) at each synapse acts on the DW synapse and updates its weight. However, since conductance of the DW synapse here takes only quantized values and is updated by write current pulses of fixed magnitude (25 $\mu$A) only (Fig. ~\ref{Fig.3}(b)), an additional quantizer circuit (Q) is present after the multiplier circuit here unlike in Bhowmik \textit{et al.}  \cite{Bhowmik_JMMM}. Design and typical output of the quantizer circuit can be found in Supplementary Material (Section 4), accompanying this paper. Despite the fact that conductance and hence weight of each synapse takes only quantized value, on-chip learning is achieved with 89 $\%$ train and 92 $\%$ test accuracy on the Fisher's Iris dataset (Table I). Test accuracy turns out to be slightly higher than train accuracy because the number of samples available in the dataset is low (100 train, 50 test), so a correct or wrong result just with respect to 1 or 2 samples changes the accuracy number by a few percent. 
\linespread{0.2}\\
Similar crossbar based FCNN is designed next with RRAM and PCM synapses, with conductance response as shown in Fig.~\ref{Fig5}. Similar accuracy for on-chip learning is achieved on Fisher's Iris dataset (Table I). However, net energy consumed in the synapses for on-chip learning is several orders of magnitude higher for RRAM/PCM synapse than DW synapse (Table I). This is expected because we already showed in Section II that energy consumed for each programming pulse that causes increase of conductance by a step (SET pulse) is orders of magnitude higher for RRAM/PCM synapse than DW synapse. Also, high energy consuming RESET pulses are still needed even though the need for decreasing conductance of a synapse is reduced by using 2 RRAM or 2 PCM per synapse \cite{PCMManan,Li} (Check Supplementary Material- Section 5). Also, training takes much longer for RRAM/ PCM synapse based FCNN compared to DW synapse based FCNN because of the need of occasional long duration RESET pulses (in microseconds). Since at each iteration (each sample in the training set) weights of all synapses need to be updated simultaneously, even if one synapse needs a RESET pulse of microsecond duration, time needed to carry out that iteration is in microseconds. Since DW synapse does not have this issue, time taken for each iteration during learning is 3 ns only (duration of each programming pulse for DW synapse in Fig. ~\ref{Fig.3}(b)) 

\section{Conclusion}
Thus in this paper we have shown through device and network level simulations that on-chip learning in DW synapse based NN circuit can consume much less time and energy than RRAM and PCM synapse based NN circuit.

\newpage

\title{\textbf{Supplementary Material}}
\author{}
\maketitle
\setcounter{section}{0}
\section{Simulation of domain wall synapse}

 The lateral dimensions of the device are taken to be 1000 nm $\times$ 50 nm. Thickness of the  ferromagnetic layer above the heavy metal layer is 1 nm.
 Thickness of heavy metal layer is 10 nm.
 
 Magnetization dynamics of the moments in this ferromagnetic layer is simulated in micromagnetic package "mumax3" with the following parameters: 
 saturation magnetization ($M_s$) = $7 \times 10^{5}$ A/m, Perpendicular Magnetic Anisotropy (PMA) constant (K) = $8 \times 10^{5}$ J/m$^3$, exchange correlation constant (A) =$1 \times 10^{-11}$ J/m and damping factor = 0.3 \cite{Emori_2}. Also, from \cite{Emori_2}, Dzyalonshinskii Moriya Interaction (DMI) is taken to be $1.2 \times 10^{-3}$ J/m$^2$ and hence DW acquires Neel type chirality.
 
 Triangular notch regions with Perpendicular Magnetic Anisotropy (PMA) constant = $9 \times 10^{5}$ J/m$^3$ are present on the edges of the simulated ferromagnetic layer in order to mimic defects, which can pin the domain wall if the driving current pulse magnitude is below a certain threshold.

Verilog-A model provided by \cite{Shimeng}, experimentally benchmarked against \cite{ExperimentalMemristor}, has been used for RRAM modeling.
Following the 1T1M (one transistor,one memristor) configuration \cite{Miao, Jiang, Li} we connect this RRAM device with a 65 nm technology node transistor (from UMC library) in Cadence Virtuoso circuit simulator. 
 
 \section{Simulation of RRAM synapse}

Scheme for conductance increase-

1. Voltage at the Top Electrode is 2V and Bottom Electrode is 0V, to enable conductance increase. When gate voltage pulses of fixed magnitude  and duration (200 ns) are applied at the gate of the transistor  , conductance increases and then saturates to a fixed value. When gate voltage pulses of fixed but larger magnitude are applied, conductance saturates to a higher final value (Fig. 6(a) of Supplementary Material).

2.  Voltage at the Top Electrode is again 2V and Bottom Electrode is again 0V, to enable conductance increase.  If gate voltage pulses of fixed duration (200 ns) but increasing magnitude are applied at the gate of the transistor, the conductance goes up linearly (Fig. 4(b) of main manuscript).

Scheme for conductance decrease-

1. Voltage at Top Electrode = 0 V and voltage at Bottom Electrode = 4.8 V, to enable conductance decrease. 
If we apply the same gate voltage pulses as in Fig. 4(b) of main manuscript in the reverse order in order to decrease the conductance of the synapse, we see that the conductance hardly decreases (Fig. 6(b) of Supplementary Material). 

2. With voltage at Top Electrode being 0 V and Bottom Electrode being 3.5 V, when a long duration (6 $\mu$s) and high magnitude (2.5 V) voltage pulse, known as RESET pulse, is applied at the gate of the transistor, conductance abruptly decreases from maximum to minimum value.

 \begin{figure}[!t]
    \centering
    {\includegraphics[width=0.4\textwidth]{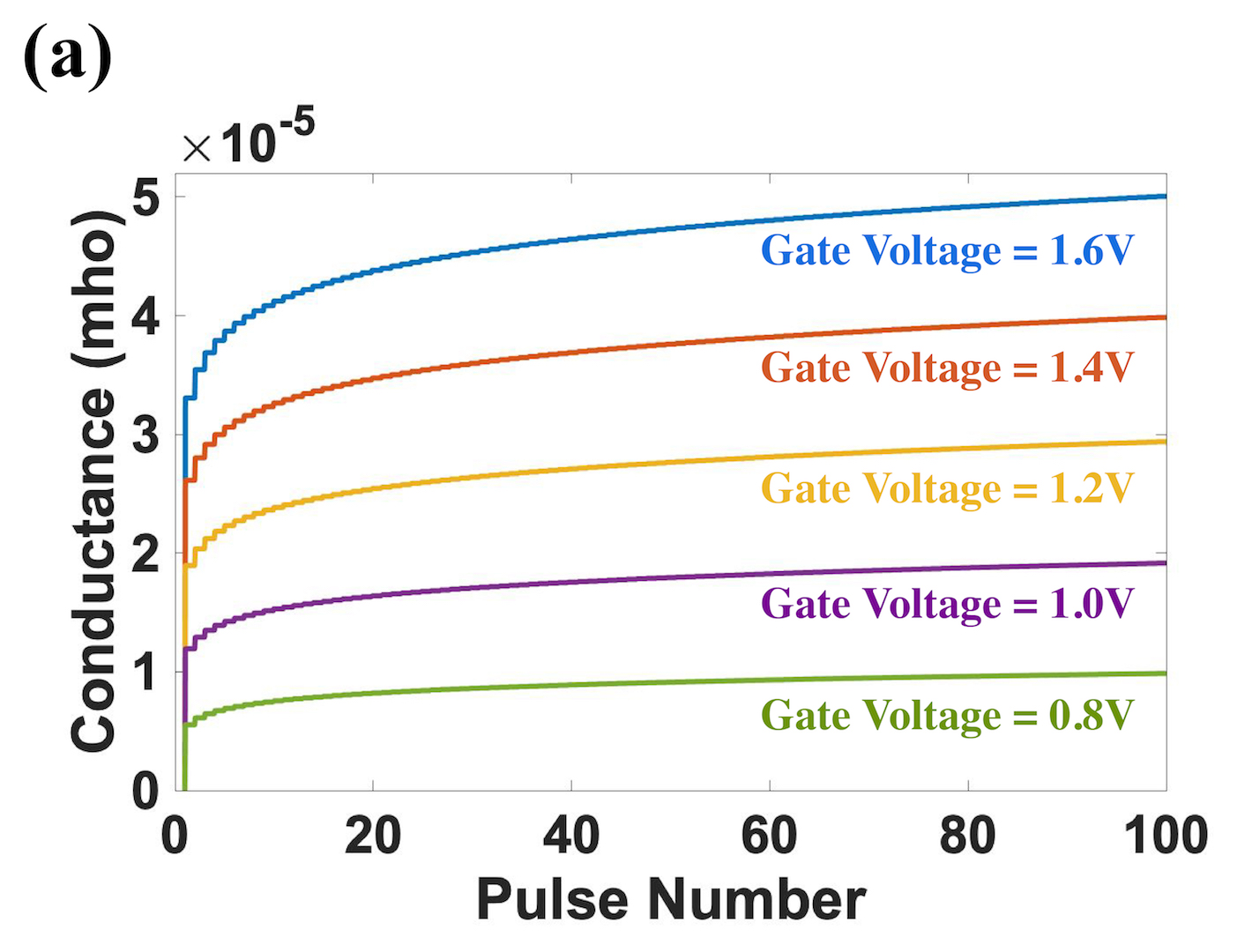}}
    {\includegraphics[width=0.45\textwidth]{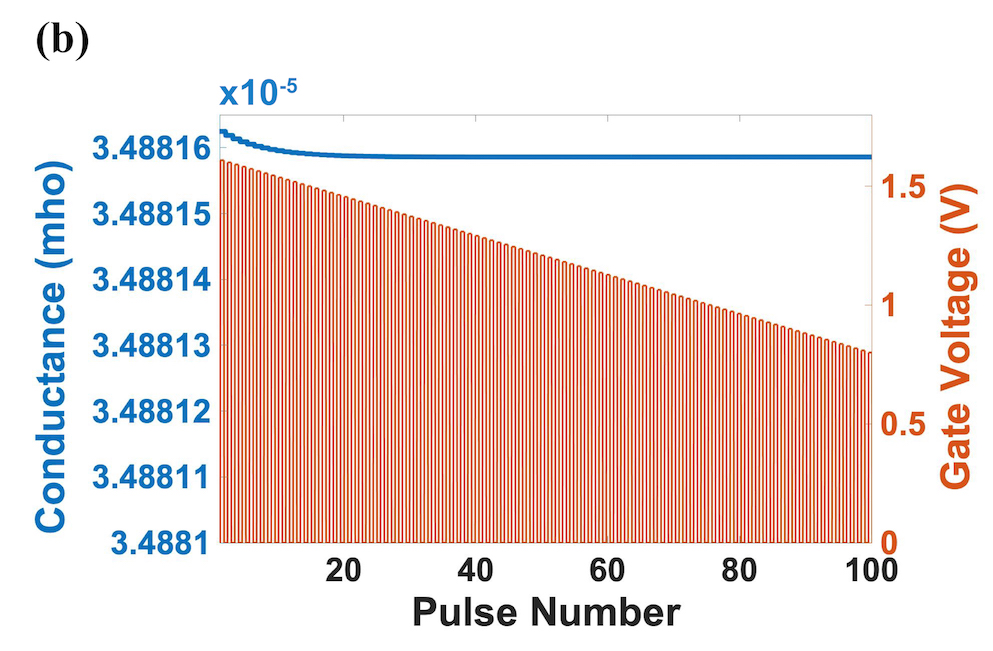}}

    \caption{(a) Non-linear increase and saturation of conductance of the 1T1M device when fixed gate voltage pulses are applied. (b)  No appreciable change in  conductance is seen by applying decreasing gate voltage pulses.  }
    
    \label{Fig1_Supplementary}
\end{figure}

 \section{Simulation of PCM synapse}

To model the conductance response of the Phase Change Memory (PCM) synapse, the model developed in \cite{Nandakumar}, based on the experiments conducted on Ge$_2$Sb$_2$Te$_5$ devices, has been used by us. By averaging the different conductance response curves generated by the model due to the stochasticity inherent in it, we have obtained the conductance response of the PCM synapse as shown in Fig. 4(c) of main manuscript.

Scheme for conductance increase-

The conductance increase characteristic is found to be more linear for PCM synapse than the case of RRAM synapse. For RRAM synapse, applying programming pulse of same magnitude led to  saturation of conductance  within first  $\approx$ 5 pulses  (Fig. 1(a) of Supplementary Material). Hence  programming pulses of linearly increasing magnitude are needed to increase the conductance linearly for a wide range of pulses and hence obtain many more conductance/ weight states (Fig. 4(b) of main manuscript). However, as observed in Fig.4(c) of main manuscript, programming current pulses (SET pulses) of magnitude 90 $\mu A$ and duration 50 ns increase the conductance of the PCM synapse fairly linearly for a larger number of pulses (~$\approx$  12) \cite{Nandakumar}. The energy associated with each such pulse is 5 pJ \cite{Nandakumar,PCMReview_AbuSebastian}.

Scheme for conductance decrease-

Conductance decrease is carried out by an abrupt RESET pulse that consumes  30pJ energy each \cite{PCMReview_AbuSebastian}, followed by a series of SET pulses (for conductance increase in small steps) much like RRAM synapse.

 \section{Quantizer Circuit for Domain Wall based Spintronic NN}
 
  \begin{figure}[!t]
    \centering
    \includegraphics[width=0.5\textwidth]{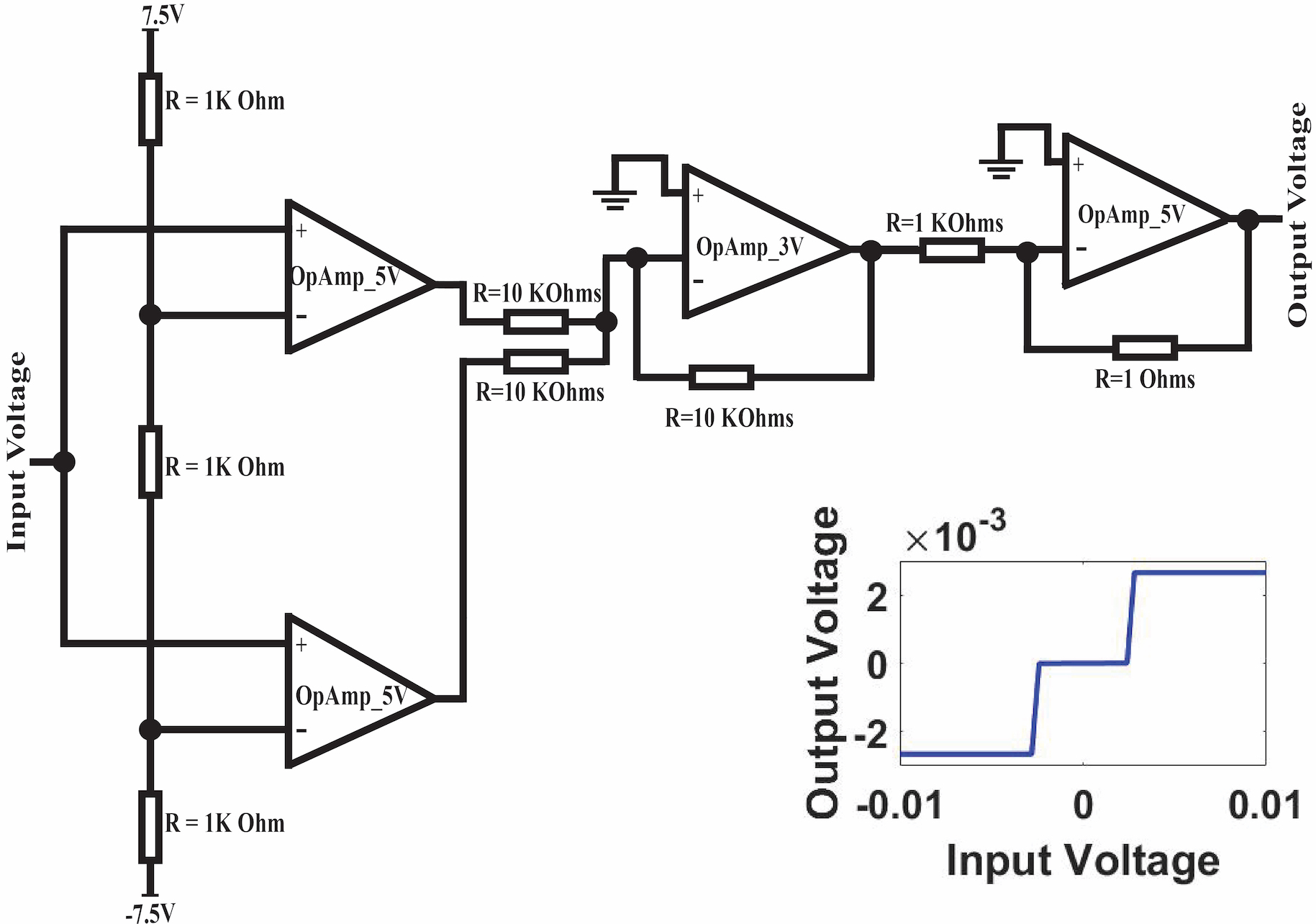}
    \caption{(a) Operational amplifier based implementation of quantizer circuit we design here. (b) Input-output characteristics of the circuit.} 
    \label{Quantizer}
\end{figure}

\begin{figure}
    \centering

    \includegraphics[width=0.49\textwidth]{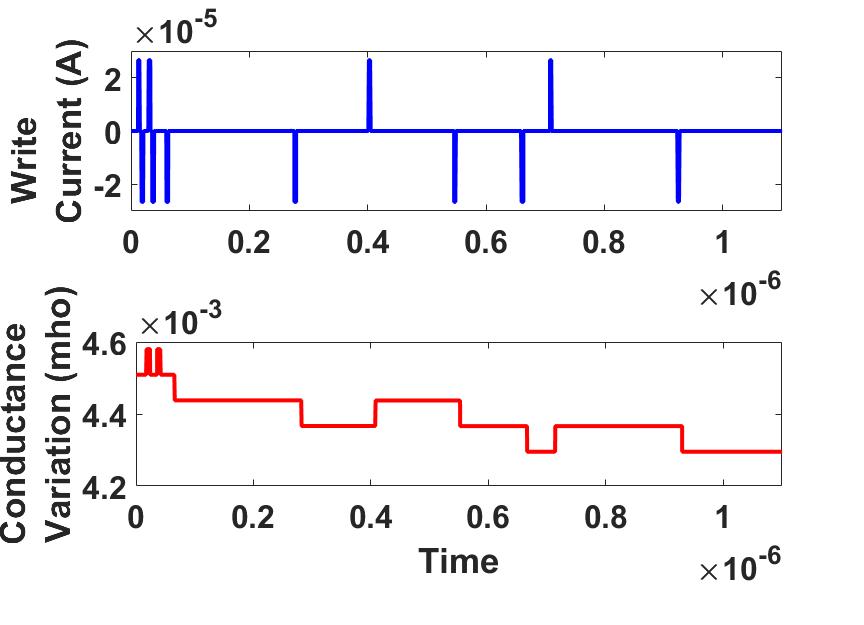}
    \caption{(a) Write current pulse (quantized) applied on a particular domain wall synapse during the course of on-chip learning of the overall domain wall synapse based neural network circuit on the Fisher's Iris dataset. (b) Corresponding change in conductance of the domain wall synapse. Since write current is quantized, conductance increases of decreases in steps of fixed magnitude. }
    \label{fig7}
\end{figure}

 To limit our "write" current to a magnitude of $25 \mu$A for either polarity, an additional "quantizer" circuit is added after the multiplier circuit which multiplies common part of weight update with the input (Fig. \ref{Quantizer} of Supplementary Material). The quantizer circuit consists of a couple of op-amps working in "Comparator" configuration, which compare the input voltage with voltages at two different points in a potential divider circuit, followed by an op-amp in "Summing amplifier" configuration which adds the output voltages of the two comparator circuits (Fig. \ref{Quantizer} of Supplementary Material).  The output of the overall quantizer circuit is hence either $\approx2.5\times10^{-3}$ V, 0 or $\approx2.5\times10^{-3}$ V.  When this output voltage is applied on the "write" terminal/ "write" path of the domain wall synapse, "write" current of three possible values ($ \approx -25\mu$A, 0 ,$ \approx 25\mu$A) flows through the domain wall synapse.  As a result conductance of the synapse goes up or down by a fixed step ($\approx 0.071\ \times 10^{-3}$ mho)  or stays unchanged (Fig. 8 of Supplementary Material).

 \section{RRAM/PCM synapse based NN}

\begin{figure}
    \centering
    \includegraphics[width=0.5\textwidth]{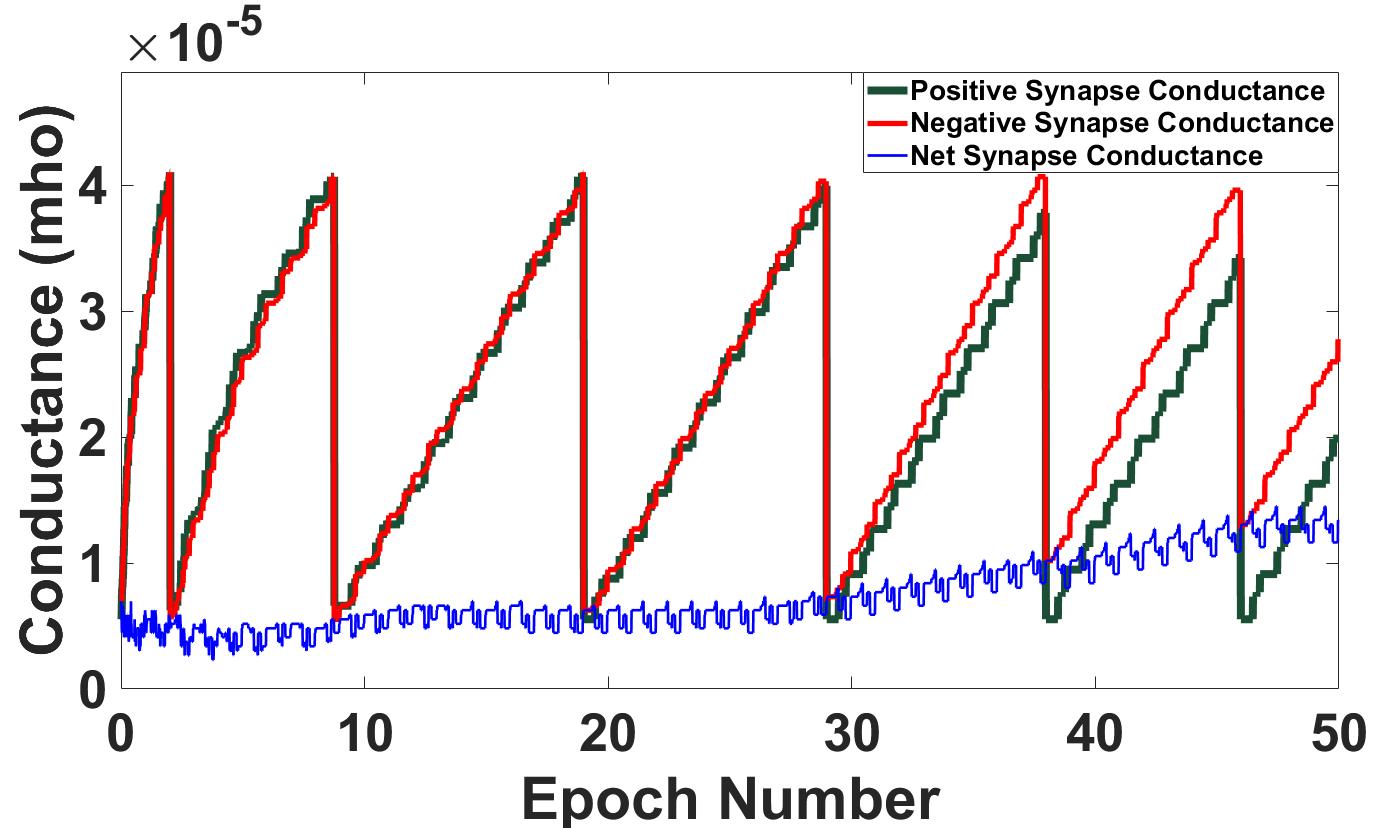}
    \caption{Conductance variation of a 2-RRAM synapse showing the need of multiple RESETs needed during on-chip training of the neural network. It can be seen that both the positive and negative synapse saturate due to conductance increase and decrease and need RESETs during training.}
    \label{RRAM_Cond_Variation}
\end{figure}

 Fig. \ref{RRAM_Cond_Variation} of Supplementary Material shows the variation of conductance values of a randomly chosen synapse in the RRAM based FCNN for 50 epochs during on-chip learning on the Fisher's Iris dataset. Under the two RRAM or PCM devices per synapse scheme \cite{Li,PCMManan} used here ,in order to increase the net conductance of the synapse, conductance of one device (positive synapse) is increased. In order to decrease the net conductance of the synapse, conductance of the other device (negative synapse) is increased. But this way, during the course of the learning, conductance of either or both synapses reaches the maximum and then Reset pulses are needed to lower the conductance to the minimum value.  Frequent use of Reset pulses increases the overall energy consumption in the scheme.

 \section {Classification accuracy of designed Domain wall, RRAM and PCM synapse based Neural Network}

 \begin{figure*}
    \centering
    {\includegraphics[width=0.5\textwidth]{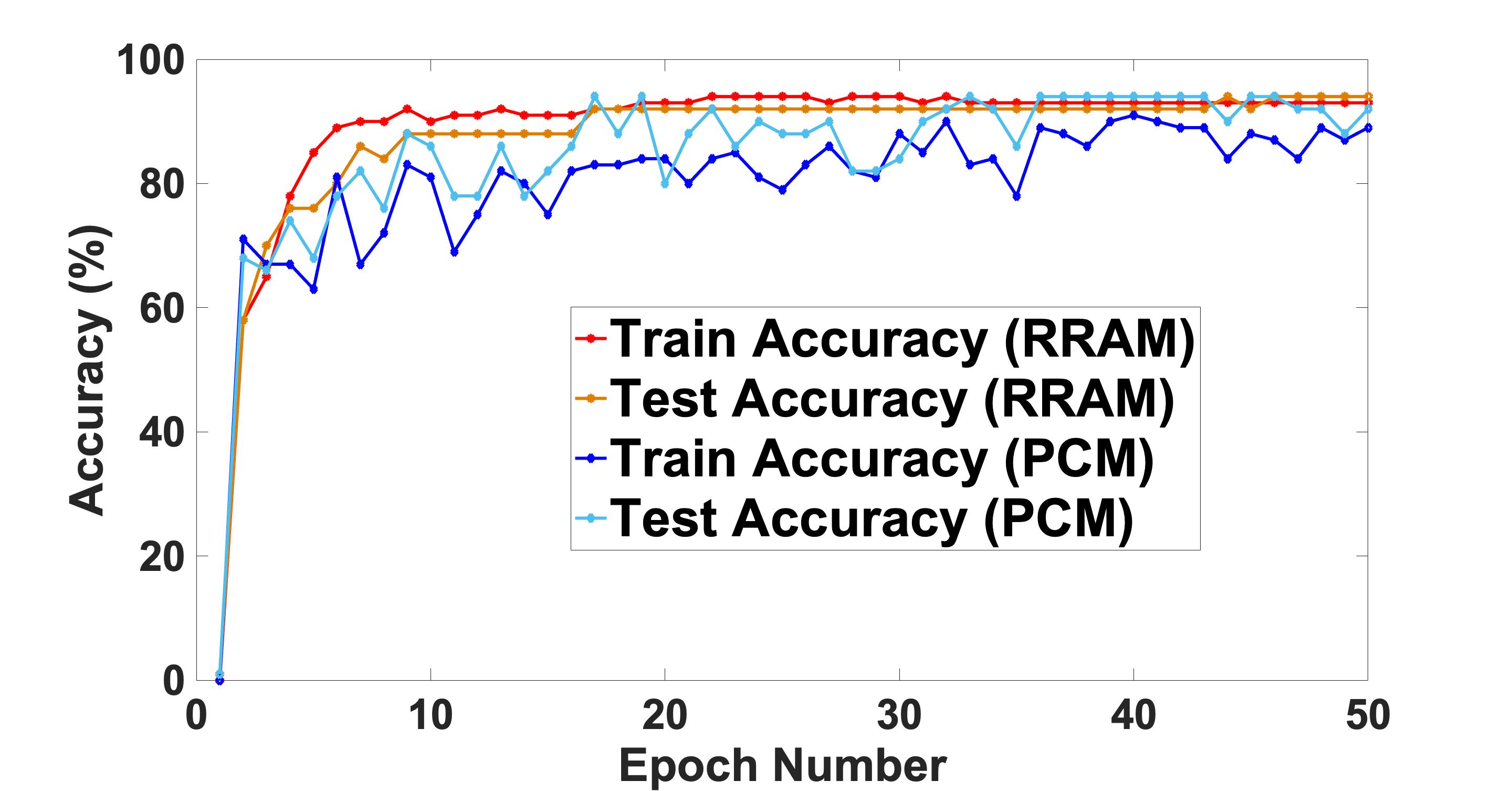}}
    {\includegraphics[width=0.455\textwidth]{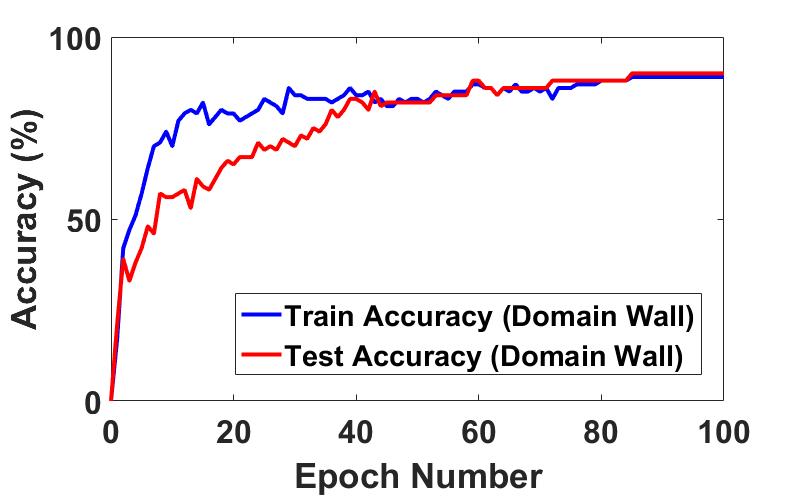}}
    \caption{(a) Train and test accuracy of RRAM and PCM synapse based neural network circuit as a function of epochs during on-chip learning on Fisher's Iris dataset. (b) Train and test accuracy of domain wall synapse based neural network circuit for the same.}
    \label{Accuracy}
\end{figure*}

Fig. \ref{Accuracy} of Supplementary Material shows the train and test accuracies of of the RRAM, PCM and spintronic (Domain wall synapse) based neural network circuits as a function of epochs for on-chip learning on the Fisher's Iris dataset.


\nocite{*}
\begin{thebibliography}{1}

 
 \bibitem{GeffBurrJPhysDReview}
H. Tsai, S. Ambrogio, P. Narayanan, R.M. Shelby, and G.W. Burr, Journal of Physics D: Applied Physics 51, 283001 (2018).

\bibitem{PCMReview_AbuSebastian}
A. Sebastian, M.L. Gallo, G.W. Burr, S. Kim, M. BrightSky et al., Journal of Applied Physics 124, 111101 (2018).

\bibitem{Kaushik_IEEEReview}
A. Sengupta, and K. Roy, IEEE Transactions on Circuits and Systems I: Regular Papers 63, 2267 (2016).


\bibitem{IBMTrueNorth}
P.A. Merolla, J.V. Arthur, R. Alvarez-Icaza, A.S. Cassidy, J. Sawada et al., Science 345, 668 (2014).

\bibitem{Loihi_Intel}
M. Davies, N. Srinivasa, T.H. Lin, G. Chinya, Y. Cao et al., IEEE Micro 38, 82 (2018).

\bibitem{GeffBurrIEDM2015}
G.W. Burr, P. Narayanan, R.M. Shelby, S. Sidler, I. Boybat, IEEE International Electron Devices Meeting, 4.4.1 (2015).

\bibitem{PCMNature}
I. Boybat, M.L. Gallo, S.R. Nandakumar, T. Moraitis, T. Parnell et al., Nature Communications 9, 1 (2018).

\bibitem{memristorNature}
Y. Li, Z. Wang, R. Midya, Q. Xia, and J.J. Yang, Journal of Physics D: Applied Physics 51, 503002 (2018).

\bibitem{PCMManan}
M. Suri, O. Bichler, D. Querlioz, O. Cueto, L. Perniola et al., IEEE International Electron Devices Meeting, 4.4.1 (2011).

\bibitem{MSuriBook}
M. Suri ed., Cognitive Systems Monographs, Springer, 2017.

\bibitem{GeffBurrTED}
 G.W. Burr, R.M. Shelby, S. Sidler, C.D. Nolfo, J. Jang et al., IEEE Transactions on Electron Devices 62, 3498 (2015).
 
\bibitem{ShimengIEDM}
P.Y. Chen, X. Peng, and S. Yu, IEEE International Electron Devices Meeting, 6.1.1 (2017).

\bibitem{RPUFrontNeuroscience}
T. Gokmen and Y. Vlasov, Frontiers in Neuroscience 10, 333 (2016).


 \bibitem{Kaushik_BioMedCircuit}
A. Sengupta, Y. Shim and K. Roy, IEEE Transactions on Biomedical Circuits and Systems 10, 1152 (2016).

\bibitem{Saxena}
U. Saxena, D. Kaushik, M. Bansal, U. Sahu and D. Bhowmik,  IEEE Transactions on Magnetics 54, 1 (2018).

\bibitem{LongYouDWSynapseExpt}
S. Zhang, S. Luo, N. Xu, Q. Zou, M. Song et al., Advanced Electronic Materials 5, 1800782 (2019).


\bibitem{Bhowmik_JMMM}
D. Bhowmik, U. Saxena, A. Dankar, A. Verma, D. Kaushik et al., Journal of Magnetism and Magnetic Materials 489, 165271 (2019).

 \bibitem{JeanAnnDomainWallNeuron1}
 N. Hassan, X. Hu, L. Jiang-Wei, W.H. Brigner, O.G. Akinola et al., Journal of Applied Physics 124, 152127 (2018).
 
  \bibitem{Parker_spintronicspiking}
 K. Yue, Y. Liu, R. K. Lake and A. C. Parker,  Science Advances 5, eaau8170 (2019).
 

 \bibitem{SkyrmionSynapse}
S. Li, W. Kang, Y. Huang, X. Zhang, Y. Zhou et al., Nanotechnology 28, 31LT01 (2017).
\bibitem{LeCun}
Y. LeCun, Y. Bengio, and G. Hinton, Nature 521, 436 (2015).
\bibitem{DWPinning1}
 S. Dutta, S. A. Siddiqui, J. A. Currivan-Incorvia, C. A. Ross and M. A. Baldo, AIP Advances 5, 127206 (2015).
 
 \bibitem{DWPinning2}
 Y. Nakatani, A. Thiaville, and J. Miltat, Nature Materials 2, 521 (2003).
 
 \bibitem{DWPinning3}
 E. Martinez, Advances in Condensed Matter Physics, 1 (2012).
 
\bibitem{Fisher}
 R. A. Fisher, Annual Eugenics 7, 179 (1936).


  
\bibitem{Emori}
  S. Emori, U. Bauer, S.M. Ahn, E. Martinez and G.S.D. Beach, Nature Materials 12, 611 (2013).
  \bibitem{Ryu}
  K.S. Ryu, L. Thomas, S.H. Yang and S. Parkin, Nature Nanotechnology 8, 527 (2013).
  \bibitem{Bhowmik}
  D. Bhowmik, M.E. Nowakowski, L. You, O. Lee, D. Keating et al., Scientific Reports 5, 11823, (2015).
 \bibitem{Miron}
I.M. Miron, T. Moore, H. Szambolics, L.D. Buda-Prejbeanu, S. Auffret et al., Nature materials 10, 419 (2011). 
 \bibitem{Emori_2}
 S. Emori, E. Martinez, K.J. Lee, H.W. Lee, U. Bauer et al., Physical Review B 90, 184427, 2014.
 
 \bibitem{Sampaio}
 J. Sampaio, V. Cros, S. Rohart, A. Thiaville and A. Fert, Nature Nanotechnology 8, 839 (2013).
 
 \bibitem{Martinez_2}
E. Martinez, S. Emori, N. Perez, L. Torres and G. S. D. Beach, Journal of Applied Physics 115, 213909 (2014).




 
 \bibitem{Sokalski}
  D. M. Bromberg, M. T. Moneck, V. M. Sokalski, J. Zhu, L. Pileggi et al., IEEE International Electron Devices Meeting, 33.1.1 (2014).

  
  
\bibitem{Zhu}
 J.-G. Zhu and C. Park, Materials Today 9, 36 (2006).
 
 \bibitem{Ikeda}
 
 S. Ikeda, K. Miura, H. Yamamoto, K. Mizunama, H.D. Gan et al., Nature Materials 9, 721 (2010). 
 
 \bibitem{mumax}
 A. Vansteenkiste, J. Leliaert, M. Dvornik, M. Helsen, F. Garcia-Sanchez et al., AIP Advances 4, 107133 (2014).

 \bibitem{Liu_Science}
  L. Liu, C.F. Pai, Y. Li, H.W. Tseng, D.C. Ralph et al., Science 336, 555 (2012).
  
  \bibitem{Liu_PRL1}
  L. Q. Liu, T. Moriyama, D. C. Ralph and R. A. Buhrman, Physical Review Letters 106, 036601 (2011).
  \bibitem{Liu_PRL2}
  L. Q. Liu, O. J. Lee, T. J. Gudmundsen, D. C. Ralph and R. A. Buhrman, Physical Review Letters 109, 096602 (2012).
 \bibitem{Shimeng}
X. Guan, S. Yu and H.S.P. Wong, IEEE Electron Device Letters 33, 1405 (2012).
 \bibitem{ExperimentalMemristor}
S. Yu, Y. Wu, and H.S.P. Wong., Applied Physics Letters 98, 103514 (2011).
 \bibitem{Miao}
F. Miao, W. Yi, I. Goldfarb, J.J. Yang, M.X. Zhang et al., ACS Nano 6, 2312 (2012).
 \bibitem{Jiang}
H. Jiang, L. Han, P. Lin, Z. Wang, M.H. Jang et al., Scientific Reports 6, 28525 (2016).
 \bibitem{Li}
C. Li, D. Belkin, Y. Li, P. Yan, M. Hu et al., Nature Communications 9, 2385 (2018).
\bibitem{Woo}
J. Woo, K. Moon, J. Song, M. Kwak, J. Park et al., IEEE Transactions on Electron Devices 63, 5064 (2016).
\bibitem{UMichNatureElectronics}
F. Cai, J.M. Correll, S.H. Lee, Y. Lim, V. Bothra et al., Nature Electronics 2, 290 (2019).
 \bibitem{Bichler}
O. Bichler, M. Suri, D. Querlioz, D. Vuillaume, B. DeSalvo et al., IEEE Transactions on Electron Devices 59, 2206 (2012).

\bibitem{Nandakumar}
S. R.Nandakumar, M. Le Gallo, I. Boybat, B. Rajendran, A. Sebastian et al., Journal of Applied Physics 124, 152135 (2018).

\bibitem{generation}
A. Sengupta, and K. Roy, Applied Physics Reviews 4, 041105 (2017).

\bibitem{Udayan}
A. Biswas, S. Prasad, S. Lashkare and U. Ganguly, arXiv preprint arXiv:1612.02233 (2016).



\end{thebibliography}
\end{document}